# What Can Student-AI Dialogues Tell Us About Students' Self-Regulated Learning? An exploratory framework

Long Zhang[1,2], Fangwei Lin[2], Weilin Wang[2,3]

South China University of Technology[1]
The University of Hong Kong[2]
Shantou University[3]

**Abstract**

The rise of Human-AI Collaborative Learning (HAICL) is shifting education toward dialogue-centric paradigms, creating an urgent need for new assessment methods. Evaluating Self-Regulated Learning (SRL) in this context presents new challenges, as the limitations of conventional approaches become more apparent. Questionnaires remain interrupted , while the utility of non-interrupted metrics like clickstream data is diminishing as more learning activity occurs within the dialogue. This study therefore investigates whether the student-AI dialogue can serve as a valid, non-interrupted data source for SRL assessment. We analyzed 421 dialogue logs from 98 university students interacting with a generative AI (GenAI) learning partner. Using large language model embeddings and clustering, we identified 22 dialogue patterns and quantified each student's interaction as a profile of alignment scores, which were analyzed against their Online Self-Regulated Learning Questionnaire (OSLQ) scores. Findings revealed a significant positive association between proactive dialogue patterns (e.g., post-class knowledge integration) and overall SRL. Conversely, reactive patterns (e.g., foundational pre-class questions) were significantly and negatively associated with overall SRL and its sub-processes. A group comparison substantiated these results, with low-SRL students showing significantly higher alignment with reactive patterns than their high-SRL counterparts. This study proposed the Dialogue-Based Human-AI Self-Regulated Learning (DHASRL) framework, a practical methodology for embedding SRL assessment directly within the HAICL dialogue to enable real-time monitoring and scaffolding of student regulation.

*Keywords:* Architectures for Educational Technology System, Evaluation Methodologies, Data Science Applications in Education

## 1. Introduction

The integration of generative artificial intelligence (GenAI) is causing a fundamental shift in educational environments, transitioning instruction from static information delivery toward dynamic, conversational learning experiences (Yan et al., 2024). This trend is particularly salient in Human-AI Collaborative Learning (HAICL), where students engage in sustained dialogues with AI agents to clarify concepts and request tailored examples (Kasneci et al., 2023). As a result, the student-AI dialogue is becoming a primary medium for cognitive engagement in the learning process (Chan & Hu, 2023).

This development presents a significant challenge to the assessment of self-regulated learning (SRL), which is defined as the proactive management of cognition, motivation, and behavior to achieve learning goals (Zimmerman, 2002). SRL is a key predictor of academic success in technology-enhanced settings (Broadbent & Poon, 2015; Zhao et al., 2025). However, the dialogue-centric nature of HAICL renders conventional methods for assessing SRL inadequate.

Historically, SRL has been measured through two main approaches. The first relies on self-report instruments such as questionnaires or structured interviews. Although these methods provide foundational data, they are interrupted (interrupting) the interactive learning process, disrupting the natural flow of learning activities, and capture students' general perceptions rather than their regulatory processes as they occur (Van Halem et al., 2020). The second approach utilizes non-interrupted trace data, such as system clickstreams, which has proven effective for measuring SRL in conventional online platforms (Fan et al., 2022; Du et al., 2023).

With advances in AI, recent research has attempted to enhance these established methods. One line of inquiry leverages GenAI to automate traditional self-report, developing AI-based structured interview systems to identify SRL behaviors (Radović et al., 2025). While this reduces the manual workload for researchers, it retains the fundamental limitation of being interrupted . Another line of research focuses on augmenting trace data analysis by integrating GenAI into learning platforms while continuing to use students' clickstream behavior to infer SRL (Ma et al., 2026; Li et al., 2025). This approach, however, does not fully account for the paradigm shift toward dialogue-centric learning. As students consolidate their activities within a single, continuous AI conversation, clickstream events become an increasingly sparse and indirect proxy for their cognitive and metacognitive processes.

The limitations of current assessment approaches underscore a critical gap: as learning becomes increasingly conversational, traditional methods often fail to provide unobtrusive yet valid monitoring of students' regulatory skills. This study contends that a robust solution lies within the dialogue data itself. Drawing on the established premise that interpersonal discourse reflects and scaffolds self-regulation (e.g., Meyer & Turner, 2002), we posit that student-AI dialogues constitute a rich, uninterrupted data stream that externalizes learners' intentions, strategies, and metacognitive states. Unlike behavioral proxies such as clickstream data (e.g., Du et al., 2023), which often lack granularity during active discourse, dialogue data captures the cognitive nuances of the interaction. By integrating clickstream logs with dialogue transcripts, researchers can bridge the "blind spots" in student tracking on GenAI-supported platforms. This multimodal synthesis provides a more holistic trace of the learning process, effectively complementing self-reports to enable high-fidelity SRL measurement and targeted interventions.

Accordingly, the primary objective of this study is to develop and validate a new assessment methodology for SRL that is suited to the specific conditions of HAICL. We investigate how patterns within student-AI dialogues can function as valid, real-time indicators of SRL. To this end, we use machine learning techniques to analyze dialogue logs, identify distinct interaction patterns, and correlate these patterns with established SRL measures. This research is guided by the following questions (RQs):

RQ1: Which dialogue patterns emerge more when students use a GenAI for learning?

RQ2: Which dialogue patterns indicate effective SRL strategies for adaptive teaching or warn SRL challenges for informing interventions?

RQ3: What are the significant differences in the dialogue patterns between the high-level and low-level SRL groups?

The contributions of this research are threefold. Theoretically, the findings map observable dialogue behaviors to foundational models of SRL (e.g., Zimmerman, 2002). Methodologically, the study validates a novel approach for dialogue-based SRL inference that is suitable for integration into adaptive systems (Li et al., 2025). Practically, this work culminates in the exploratory Dialogue-Based Human-AI Self-Regulated Learning (DHASRL) framework, which extends contemporary models of self-regulation for the HAICL context (e.g., Järvelä et al., 2023). This framework provides an operational basis for monitoring and scaffolding student regulation in real-time, a necessary component for designing learning environments that are truly responsive to student intent.

## 2. Literature Review

### 2.1. Self-Regulated Learning

SRL is a multifaceted construct describing how learners proactively and intentionally direct their efforts to manage their own learning processes. Rather than passively receiving information, self-regulated learners actively engage metacognitively, motivationally, and behaviorally in their learning. Process-oriented models are particularly relevant for analyzing SRL as it unfolds over time. Zimmerman's (2002) cyclical model is a highly influential example, conceptualizing regulation across three phases: forethought (planning before a task), performance (monitoring during a task), and self-reflection (evaluating after a task). Pintrich's (2004) framework similarly embodies this cyclical nature, with added emphasis on the role of affect and motivation. Winne and Hadwin's (1998) model outlines four stages of information processing, highlighting the internal cognitive conditions and standards that guide learning. A common thread in these established models is their primary focus on the individual learner. However, as learning has become increasingly collaborative and technology-mediated during AI integrating, SRL theory has expanded to account for the social contexts in which regulation occurs.

The evolution of SRL theory has introduced concepts such as co-regulated learning (CoRL) and socially shared regulation of learning (SSRL) (Hadwin et al., 2017). CoRL, particularly relevant here, describes the process where a more knowledgeable other provides scaffolding to support a learner's regulatory processes, which is gradually withdrawn as the learner internalizes those skills. This co-regulatory activity is fundamentally dialogic, involving prompts, feedback, and strategic guidance. Recent work has modeled this interplay between humans and AI, with proposing a framework for Human-AI Shared Regulation in Learning (HASRL, Järvelä et al., 2023). In this model, the AI detects regulatory "Traces" and "Signals" from the learner's activity, diagnoses their regulatory state, and acts by providing support. This framework positions dialogue as a key data source for diagnosis (e.g., Edwards et al., 2025); however, the empirical link between specific dialogue learning patterns and the learner's SRL state remains unestablished. Framed as dialogic co-regulation, HAICL offers a novel context for SRL research.

**2.2. Measuring SRL in Technology-Enhanced Learning Environments**

SRL measurement has advanced with technology, balancing theoretical depth and practicality. The most traditional and widely used method involves self-report instruments, such as the Online Self-Regulated Learning Questionnaire (OSLQ) (Barnard et al., 2009), which we use in this study as a baseline measure. These questionnaires ask learners to rate their agreement with statements about their use of various cognitive and metacognitive strategies. Their primary strengths are their ease of administration and their foundation in validated psychological constructs. However, they are also subject to significant limitations, including recall bias (learners may not accurately remember their past behaviors), social desirability bias (learners may report what they think is expected), and their static, aggregated nature. A self-report score provides a global trait-like assessment but cannot capture the dynamic, in-the-moment processes of regulation as they unfold during a specific learning task (Van Halem et al., 2020).

The proliferation of online learning environments has generated a novel data source: trace data of digital footprints, also known as clickstream data. Recent researches have utilized these logs to infer SRL processes from behaviors such as resource access frequency, forum participation, and video-viewing patterns (Du et al., 2023; Fan et al., 2022). The strength of this approach lies in its ability to capture actual behavior at scale and over time, thus addressing some limitations of self-report measures (Van Halem et al., 2020). However, clickstream data provides an incomplete picture. While we can observe what a student did (e.g., viewed a page three times), we often lack insight into why they did it. As Fan et al. (2022) argue, articulating students' thought processes is essential for a comprehensive understanding. Consequently, the era of trace data analysis has revealed its inherent limitations, underscoring the need for richer data sources to elucidate the strategic intent underlying student actions.

The integration of GenAI into learning environments, particularly within HAICL contexts, has introduced a transformative data source: student-AI dialogues. While prior research has established that peer-to-peer interactions both reflect and scaffold SRL (e.g., Meyer & Turner, 2002), the rapid evolution of GenAI potentially suggests that human-AI dialogues are equally capable of manifesting and supporting these regulatory processes. These dialogues offer unique windows into students' cognitive and metacognitive trajectories, providing a critical triangulation point for self-report measures—such as interviews and questionnaires—and objective clickstream data (Li et al., 2025; Radović et al., 2025; Van Halem et al., 2020). By capturing students' verbalized thought processes during interactions with AI agents, these dialogues hold potential for a more comprehensive assessment of SRL competencies. Our study aims to provide the foundational data required to empirically link dialogue learning patterns with underlying SRL competence, thereby informing the diagnostic component of models such as Järvelä et al.'s (2023). By validating dialogue as a meaningful indicator of HAI co-regulated learning, we have exploringly expanded the operationalized framework of the HASRL in the context of human-AI dialogue learning (such as DHASRL).

## 3. Methods

### 3.1. Participants

The study sample comprised 106 second-year undergraduate students (67% female, 33% male; age range 19-20) from a university. All participants were enrolled in a mandatory introductory statistics course as a required component of their undergraduate finance program. This specific cohort was selected to ensure a relatively homogeneous baseline of prior domain knowledge and academic motivation. To enhance the ecological validity of the study, recruitment was conducted during the middle weeks of the semester. This timing ensured that students possessed sufficient foundational knowledge of key concepts from their coursework but were not under the immediate cognitive and motivational pressures associated with final examinations.

Participation was strictly voluntary. They were explicitly informed at the outset that their interactions and performance within the experimental setting would have no bearing on their official course grades. No monetary or other academic incentives were offered. The research protocol received full approval from the university's Institutional Review Board.

### 3.2. Instruments

Students' SRL capacities were measured using the Online Self-Regulated Learning Questionnaire (OSLQ; Barnard et al., 2009), consisting of 24 items across six subscales: goal setting (GS), environment structuring (ES), task strategies (TS), time management (TM), help-seeking (HS), and self-evaluation (SE). The overall Cronbach's alpha was 0.914, confirming strong internal consistency. Its measurement in an AI-supported environment still ensures good instrument's validity (Zhang & Chen, in press).

Furthermore, for the dialogue patterns, we directly mined them from the student-AI dialogue logs. Therefore, we directly collected their dialogue logs.

**3.3. Experiment and Data Collection procedure**

Given the exploratory nature of this study, we adopted an observational approach to collect dialogue data from students utilizing AI for self-directed learning. Our objective was to identify conversational patterns that correlate with students' SRL capabilities (measured by OSLQ).

The experiment was conducted within a bespoke HAICL system (supported by doubao v1.6) developed for this study. A core design principle of the system was pedagogical transparency. When a student submitted a query, the system's interface displayed the final, complete prompt that was constructed via a predefined template engineered to optimize response quality. Furthermore, the system's output presented not only the final answer but also the intermediate reasoning steps generated by the large language model. This design choice provided students with explicit insight into the AI's generation process.

The study was executed over a four-week period within an authentic undergraduate statistics course. Prior to the start of the experiment, all participating students received a training session on how to use the HAICL system effectively. The experimental task required students to use the system for two distinct purposes each week: first, to prepare for the upcoming lecture (pre-class task), and second, to review course material and plan subsequent learning activities after the lecture (post-class task). At the conclusion of the four-week period, students were instructed to submit their complete dialogue logs with the AI. Immediately following the submission, each student completed the OSLQ for SRL assessment. After accounting for 8 students did not take the OSLQ test., the final dataset available for analysis included 421 dialogue logs and 98 completed OSLQ assessments.

**3.4. Data Analysis**

The statistical analysis was performed on a set of 22 alignment scores computed for each student, which were generated using an approach based on large language model embeddings and clustering. Instead of reducing a conversation to a single dominant category, this method quantifies a student's dialogue as a distribution of alignment across all 22 patterns. It preserves the nuanced characteristics of the original conversation, yielding a comprehensive behavioral profile for each learner that reflects their unique blend of interactional strategies. The technical procedure and metrics (e.g. Silhouette Score) for calculating these alignment scores is detailed in Appendix A.

To address the RQ1, which patterns were most representative of the conversations, we analyzed the distribution of these alignment scores. Specifically, we used descriptive statistics, reporting the mean and standard deviation for each pattern's scores to characterize its overall prominence and variability across the student sample.

To answer RQ2, which explores the relationship between dialogue patterns and self-regulated learning, we conducted a correlational analysis. We employed Spearman's rank correlation to compute the associations between each student's 22 dialogue pattern alignment scores and their scores on the overall SRL measure, as well as its six sub-dimensions (GS, ES, TS, TM, HS, and SE). Given the large number of correlations calculated (22 patterns × 7 SRL scores), the Benjamini-Hochberg False Discovery Rate (FDR) procedure was applied to adjust p-values and control for the increased risk of Type I errors inherent in multiple comparisons.

To address RQ3 regarding the differences in dialogue patterns between learner groups, we partitioned the participants into high-SRL and low-SRL groups using a median split of their total SRL scores. While we acknowledge the methodological debates surrounding the dichotomization of continuous variables (MacCallum et al., 2002), this approach was adopted to facilitate the interpretability of distinct learner profiles and is considered a statistically robust procedure when the objective is to contrast group-based behavioral patterns (Iacobucci et al., 2015). We then conducted a series of Mann-Whitney U tests, a non-parametric procedure, to compare the distributions of alignment scores for each of the 22 dialogue patterns between these two independent groups. To mitigate the risk of Type I errors arising from multiple comparisons, FDR correction was applied to the resulting p-values across all 22 tests.

Finally, through the answers to RQ1-RQ3, we proposed an evidence-based exploratory framework, DHASRL.

## 4. Results

Data preprocessing produced a total of 22 discrete conversational clusters: five for pre-class student questions (PreQ), nine for post-class student questions (PostQ), two for pre-class AI responses (PreR), and six for post-class AI responses (PostR). The final interpretations, which were used as feature variables in the subsequent analyses, are presented in Table 1. The identified clusters represent a wide range of conversational behaviors, from students asking about surface-level facts (PreQ_cluster0) to engaging in high-level metacognitive reflection (PostQ_cluster8), and from AI responses providing foundational concepts (PreR_cluster0) to offering deep methodological elaborations (PostR_cluster5). To address RQ1, we generated violin-box plot to visualize the distributions of the 22 dialogue patterns (Fig. 1).

**Table 1.**

*Clusters and Interpretation based on large language model embeddings and clustering (refer detail for Appendix A)*

| Cluster ID | Interpretation |
|---|---|
| *Student Questions* | |
| PreQ_cluster0 | **Surface-Level Prior Knowledge Gaps.** Questions revealing a lack of basic, prerequisite factual knowledge (e.g., definitions of core statistical terms). These indicate potential deficits in prior learning or preparation. |
| PreQ_cluster1 | **Foundational Conceptual Difficulties.** Questions expressing significant difficulty understanding fundamental course concepts and formulas. These signal a systemic learning obstacle and a potential failure in prior self-evaluation and mastery of prerequisite skills. |
| PreQ_cluster2 | **External Learning Environment Obstacles.** Dialogue focused on external factors that impede learning, such as equipment problems, environmental noise, and other distractions. This externalizes an awareness of factors related to Environment Structuring. |
| PreQ_cluster3 | **Internal Learning Distractions & Coping Strategies.** Questions centered on internal states of distraction and proactive strategies for managing them. This reflects a metacognitive awareness of Effort Regulation and attention control. |
| PreQ_cluster4 | **Deep Learning & Skill Transfer.** Inquiries focused on summarizing learning, exploring problem-solving methodologies, and applying knowledge to |
| | practical scenarios. This represents a need for deeper processing and reflects cognitive strategies related to Elaboration and Critical Thinking. |
| PostQ_cluster0 | **Post-Instruction Conceptual Gaps.** Questions indicating a continued lack of understanding of foundational course content *after* instruction, similar in nature to PreQ_cluster1. This suggests a failure to comprehend the material during the performance phase. |
| PostQ_cluster1 | **Post-Instruction Strategy & Focus.** Dialogue focused on strategic adjustments to learning environment and attention *after* a learning session, similar in nature to PreQ_cluster3. |
| PostQ_cluster2 | **Targeted Content Clarification.** Focused questions on specific, core lecture topics (e.g., applying a specific function type), indicating an attempt to resolve precise points of confusion after initial instruction. |
| PostQ_cluster3 | **Environment & Distraction Management.** Dialogue reflecting on the learning environment, sources of distraction, and coping strategies. This signifies metacognitive activity related to Environment Structuring. |
| PostQ_cluster4 | **Time and Environment Management Planning.** Dialogue explicitly discussing the planning and structuring of future study, such as scheduling regular evening study times and securing necessary equipment. This directly externalizes Time Management and Environment Structuring strategies. |
| PostQ_cluster5 | **Systematic Knowledge Integration.** Questions covering a broad range of topics from basic concepts to computational methods and applications, suggesting a systematic effort to organize and integrate knowledge post-instruction. |
| PostQ_cluster6 | **Reflection on Environment and Attention.** Dialogue describing and reflecting on the chosen study place, time, and personal state of attention, indicating metacognitive monitoring of one's study environment. |
| PostQ_cluster7 | **Strategies for Maintaining Focus.** Questions explicitly about how to maintain concentration and deal with distractions, indicating a search for adaptive strategies related to Effort Regulation. |
| PostQ_cluster8 | **Metacognitive Reflection and Summary.** Dialogue focused on reviewing the learning process, summarizing acquired knowledge, and evaluating one's own understanding. This is a direct indicator of Self-Evaluation and reflective processes. |
| *AI Responses* | |
| PreR_cluster0 | **Foundational Concept Provision.** AI responses that introduce and explain fundamental statistical concepts and formulas (e.g., hypothesis testing, t-test) in preparation for an upcoming lecture. These are typically triggered by students' lack of prerequisite knowledge. |
| PreR_cluster1 | **Advanced Conceptual Preparation.** AI responses that explain more advanced statistical methods and their application logic, often using analogies. These are triggered by students seeking to proactively understand the 'why' and 'how' of complex topics before the lecture. |
| PostR_cluster0 | **Remedial Review of Basic Calculations.** AI responses that review and reinforce the most elementary statistical concepts and calculation procedures |

| | (e.g., mean, median, mode) after a lecture. This serves a remedial function for students who missed or failed to grasp foundational content during class. |
|---|---|
| PostR_cluster1 | **Extracurricular Conceptual Extension.** AI responses that delve into advanced topics (e.g., central limit theorem, sampling inference) that are typically beyond the immediate scope of the curriculum. These are triggered by student curiosity and a desire for deeper, cross-disciplinary knowledge. |
| PostR_cluster2 | **Targeted Formula Deconstruction.** AI responses that provide a granular, step-by-step breakdown of a specific statistical formula discussed in class, explaining the logic and meaning of each component. This serves to remediate a precise point of confusion about a procedure. |
| PostR_cluster3 | **Comparative Method Selection.** AI responses that help students compare and contrast different statistical methods (e.g., parametric vs. non-parametric tests) based on their application scenarios. This supports the cognitive task of organizing knowledge and making strategic choices. |
| PostR_cluster4 | **Guidance from Concept to Application.** AI provided a comprehensive explanation of the concept, principles, and application scenarios of the vector method (e.g., PCA, matrix diagonalization). Due to the concept being difficult to understand, the AI's response covered the transition from theory to actual research application. |
| PostR_cluster5 | **Deep Methodological Elaboration.** AI responses that provide a comprehensive, deep-dive explanation of a core course methodology (like the t-test), connecting theory, formula, and practical application through analogies and examples. This fosters an integrated, applied understanding post-instruction. |

**Figure. 1**

*Violin-box Plot of 22 dialogue patterns' distribution*

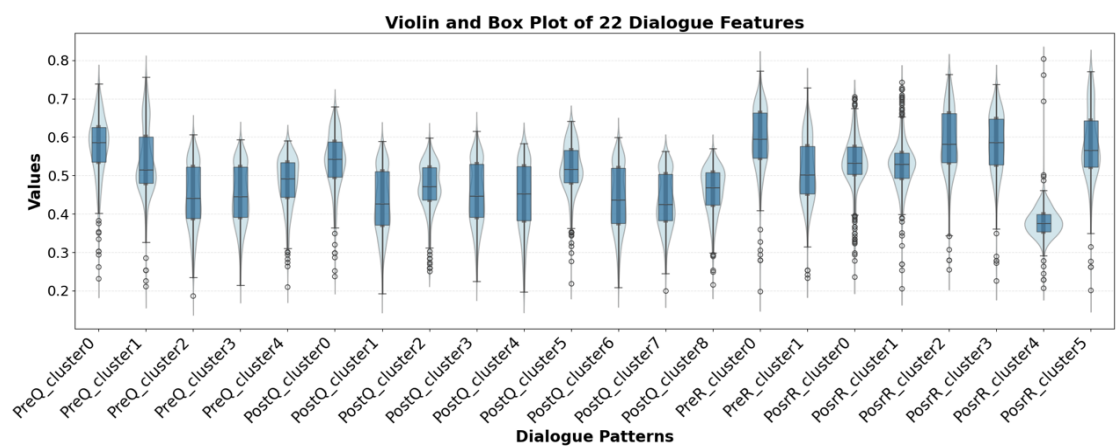

Figure 1 presents the distribution of student dialogues across the 22 identified patterns. The findings reveal a non-uniform distribution, indicating that certain dialogue patterns were more prominent than others in the student-AI conversations. On average, student dialogues showed the strongest alignment with patterns involving AI-provided scaffolding, particularly the provision of foundational concepts before a lecture (PreR_cluster0, M = .596) and the deconstruction of specific formulas after a lecture (PostR_cluster2, M = .589). This indicates that, overall, the conversations were most reflective of interactions centered on core content and procedural knowledge. In contrast, the pattern reflecting AI guidance on applying concepts to research contexts (PostR_cluster4, M = .380) showed the lowest mean alignment, suggesting such advanced, application-oriented dialogues were the least prominent feature of the dataset.

An examination of the standard deviations reveals substantial differences in the consistency of these patterns across learners. The widest variation was observed for patterns related to deep methodological elaborations by the AI (PostR_cluster5, SD = .093) and student expressions of foundational conceptual difficulties (PreQ_cluster1, SD = .092). This high degree of dispersion suggests that the extent to which students struggled with fundamental concepts and required in-depth support varied considerably. Conversely, dialogues concerning the application of concepts to research (PostR_cluster4, SD = .050) not only had the lowest mean alignment but also the smallest variation, indicating that such conversations were consistently uncommon among all students. The range between the minimum and maximum values for most clusters further indicates the presence of outliers, reflecting the individualized nature of students' dialogue pathways.

**4.2. Dialogue Patterns Associated with SRL**

The correlation analysis addressed the RQ2 by identifying dialogue patterns positively and negatively associated with SRL, revealing actionable insights for adaptive pedagogy. For instance, proactive patterns (e.g., post-class integrative questions) are modestly associated with stronger SRL strategies that educators can leverage to scaffold reflection and goal-setting in HAICL environments, while reactive patterns (e.g., foundational queries triggering detailed AI responses) are modestly associated with SRL challenges, suggesting the need for targeted interventions to build student autonomy. These associational results are comprehensively summarized in the heat-table visualization of Table 2 below.

**Table 2.**

*Dialogue Learning Patterns Associated with SRL*

| Cluster ID | GS | ES | TS | TM | HS | SE | SRL |
| --- | --- | --- | --- | --- | --- | --- | --- |
| PreQ_cluster0 | 0.193* | -0.092 | 0.079 | 0.181* | -0.011 | -0.002 | 0.055 |
| PreQ_cluster1 | -0.075 | -0.137* | -0.091 | -0.062 | -0.168* | -0.193* | -0.190* |
| PreQ_cluster2 | 0.193* | -0.017 | 0.021 | 0.075 | 0.065 | -0.040 | 0.054 |
| PreQ_cluster3 | 0.192* | 0.004 | 0.017 | 0.069 | 0.076 | -0.046 | 0.058 |
| PreQ_cluster4 | 0.168* | -0.047 | 0.025 | 0.067 | 0.028 | -0.074 | 0.022 |
| PostQ_cluster0 | 0.003 | -0.139* | -0.066 | -0.030 | -0.130* | -0.165* | -0.142* |
| PostQ_cluster1 | 0.191* | -0.024 | 0.015 | 0.075 | 0.065 | -0.044 | 0.049 |
| PostQ_cluster2 | 0.222* | -0.017 | 0.084 | 0.141* | 0.070 | 0.018 | 0.103 |
| PostQ_cluster3 | 0.192* | -0.027 | 0.017 | 0.078 | 0.057 | -0.046 | 0.047 |
| PostQ_cluster4 | 0.175* | -0.040 | 0.001 | 0.076 | 0.059 | -0.055 | 0.031 |
| PostQ_cluster5 | 0.293* | 0.009 | 0.147* | 0.204* | 0.061 | 0.050 | 0.164* |
| PostQ_cluster6 | 0.202* | -0.023 | 0.025 | 0.071 | 0.048 | -0.046 | 0.049 |
| PostQ_cluster7 | 0.203* | -0.016 | 0.021 | 0.081 | 0.068 | -0.046 | 0.057 |
| PostQ_cluster8 | 0.157* | -0.047 | 0.010 | 0.056 | 0.038 | -0.079 | 0.020 |
| PreR_cluster0 | -0.029 | -0.092 | -0.066 | 0.002 | -0.161* | -0.173* | -0.146* |
| PreR_cluster1 | -0.148* | 0.008 | -0.124 | -0.106 | -0.167* | -0.198* | -0.191* |
| PostR_cluster0 | 0.174* | 0.013 | 0.092 | 0.168* | 0.075 | 0.063 | 0.134* |
| PostR_cluster1 | 0.105 | 0.052 | 0.048 | 0.086 | -0.019 | -0.051 | 0.039 |
| PostR_cluster2 | -0.056 | -0.056 | -0.090 | -0.043 | -0.196* | -0.182* | -0.169* |
| PostR_cluster3 | -0.047 | -0.066 | -0.054 | -0.022 | -0.150* | -0.172* | -0.140* |
| PostR_cluster4 | -0.288* | -0.063 | -0.129* | -0.151* | -0.109 | -0.181* | -0.220* |
| PostR_cluster5 | -0.140* | -0.147* | -0.113 | -0.081 | -0.203* | -0.229* | -0.230* |

*Note.* * means p < 0.05 with FDR correction. Rows represent dialogue clusters, columns represent SRL variables, cell colors indicate r-value magnitude [red for negative, green for positive]. GS means goal setting; ES means environment structuring; TS means task strategies; TM means time-management; HS mean help-seeking; SE means self-evaluation; SRL means the total of all six sub-dimensions of OSLQ.

As outlined in Table 2, the correlational analysis revealed widespread significant associations between dialogue patterns and SRL. Of the 22 identified dialogue patterns, 21 patterns (95.5%) demonstrated at least one significant correlation with either the overall SRL score or one of its six sub-dimensions (Goal Setting, Environment Structuring, Task Strategies, Time Management, Help-Seeking, Self-Evaluation), with only one pattern (PostR_cluster1) showing no significant associations. Furthermore, these correlations spanned all seven SRL variables: Goal Setting showed significant correlations with 14 dialogue patterns, Self-Evaluation with 11 patterns, Help-Seeking with 10 patterns, Time Management with 5 patterns, Environment Structuring with 3 patterns, Task Strategies with 2 patterns, and overall SRL with 10 patterns. The correlation coefficients ranged from r = -.288 (PostR_cluster4 with Goal Setting) to r = .293 (PostQ_cluster5 with Goal Setting), indicating both negative and moderate positive associations.

Examining the directionality of these associations, 13 dialogue patterns exhibited exclusively positive significant correlations (no significant negative associations), while 8 patterns exhibited exclusively negative significant correlations (no significant positive associations). Among the positively associated patterns, Goal Setting emerged as the most frequently correlated sub-dimension, showing significant positive correlations with 12 of these 13 patterns (r = .157 to .293). Two patterns were significantly associated with overall SRL: PostQ_cluster5 (r = .164) and PostR_cluster0 (r = .134). Time Management showed significant positive correlations with 4 patterns (r = .141 to .204), and Task Strategies with 1 pattern (r = .147). Among the negatively associated patterns, Self-Evaluation and Help-Seeking were the most frequently correlated sub-dimensions, each showing significant negative correlations with all 8 patterns (Self-Evaluation: r = -.173 to -.229; Help-Seeking: r = -.130 to -.203). Overall SRL showed significant negative correlations with all 8 patterns as well (r = -.140 to -.230). Additionally, Goal Setting showed negative correlations with 2 patterns, Environment Structuring with 2 patterns, Task Strategies with 1 pattern, and Time Management with 1 pattern.

Notably, the strongest positive correlation was observed between PostQ_cluster5 (Systematic Knowledge Integration) and Goal Setting (r = .293), while the strongest negative correlation was between PostR_cluster4 (Guidance from Concept to Application) and Goal Setting (r = -.288). For overall SRL, the strongest positive association was with PostQ_cluster5 (r = .164), whereas the strongest negative associations were with PostR_cluster5 (Deep Methodological Elaboration, r = -.230) and PostR_cluster4 (Guidance from Concept to Application, r = -.220). Across all patterns, no dialogue pattern exhibited both significant positive and significant negative correlations simultaneously, indicating that each pattern was consistently associated with either beneficial or detrimental SRL profiles.

**4.3. Dialogue Patterns Difference Between High/Low SRL Level Groups**

**Table 3.**

*Dialogue Patterns Difference Between High/Low SRL Level Groups By Mann-Whitney U Tests*

| Cluster ID | High SRL Group | Low SRL Group | U | p |
| --- | --- | --- | --- | --- |

|               | Mean(SD)    | Median | Mean(SD)    | Median |           |       |
|---------------|-------------|--------|-------------|--------|-----------|-------|
| PreQ_cluster1 | 0.520(0.09) | 0.508  | 0.555(0.09) | 0.521  | 13943.500 | 0.009 |
| PostQ_cluster0| 0.525(0.08) | 0.535  | 0.549(0.07) | 0.550  | 14732.500 | 0.042 |
| PreR_cluster0 | 0.583(0.09) | 0.592  | 0.608(0.08) | 0.602  | 14727.500 | 0.042 |
| PreR_cluster1 | 0.495(0.08) | 0.487  | 0.527(0.08) | 0.523  | 13711.500 | 0.006 |
| PostR_cluster2| 0.575(0.09) | 0.570  | 0.603(0.08) | 0.605  | 14407.500 | 0.029 |
| PostR_cluster3| 0.570(0.09) | 0.579  | 0.595(0.08) | 0.594  | 14507.500 | 0.031 |
| PostR_cluster5| 0.559(0.09) | 0.549  | 0.598(0.09) | 0.578  | 13430.500 | 0.004 |

*Note*: Significance p-value calculation employed FDR correction.

To investigate the differences in dialogue patterns between high- and low-SRL groups, a series of Mann-Whitney U tests were conducted (Tab. 3). The results, as detailed in the table, reveal that students in the low-SRL group exhibited significantly more frequent questioning related to fundamental learning challenges. Specifically, they were more likely to ask about foundational conceptual difficulties before instruction (PreQ_cluster1; M_low = .555 vs. M_high = .520, p = .009) and to express post-instruction conceptual gaps (PostQ_cluster0; M_low = .549 vs. M_high = .525, p = .042). This indicates that learners with lower SRL proficiency are characterized by a persistent struggle with core course material, a pattern that manifests in their dialogues both before and after formal teaching sessions. These findings provide direct evidence that specific, student-initiated questioning patterns can effectively differentiate between SRL levels.

Furthermore, the analysis demonstrates that the AI's responses, as triggered by the two groups, were also significantly different. The low-SRL group prompted the AI to provide a greater frequency of scaffolding-intensive responses. This included both preparatory explanations, such as foundational concept provision (PreR_cluster0) and advanced conceptual preparation (PreR_cluster1), as well as remedial support after lectures, such as targeted formula deconstruction (PostR_cluster2). Most notably, the low-SRL group required significantly more deep methodological elaborations (PostR_cluster5; M_low = .598 vs. M_high = .559, p = .004). This pattern of consistently requiring more extensive and varied support from the AI, from basic review to in-depth explanation, serves as a robust digital footprint of learners facing SRL challenges, thereby confirming that interaction data can be a powerful tool for tracking SRL status.

## 5. Discussion

### 5.1. Distribution of Dialogue Patterns

An initial examination of the descriptive statistics provides the foundational context for this study's primary objective. The analysis reveals two key findings. First, the prevalence of dialogues centered on foundational concepts (e.g., PreR_cluster0) and the relative scarcity of advanced application-oriented dialogues (PostR_cluster4) ground the investigation in the reality of this specific learning context, where the primary challenge appears to be content mastery. Second, and more central to the goal of identifying SRL indicators, is the high degree of variation observed in patterns associated with conceptual difficulty (PreQ_cluster1) and intensive AI support (PostR_cluster5). This heterogeneity is a critical prerequisite for these patterns to serve as valid indicators; if all learners exhibited similar dialogue behaviors, it would not be possible to differentiate between SRL levels. The observed variance confirms that sufficient behavioral differences exist to warrant the subsequent investigation into their relationship with self-regulation.

### 5.2. Dialogue Patterns Associated with SRL

5.2.1. Proactive Regulation Patterns

The analysis revealed two conversational behaviors that demonstrated a significant, positive association with the overall SRL score. Most prominently, students' Systematic Knowledge Integration (PostQ_cluster5) after instruction correlated significantly with higher overall SRL ($r = 0.164$). Theoretically, this pattern embodies the high-level cognitive organization central to Zimmerman's (2002) performance and self-reflection phases, where learners actively synthesize disparate concepts to construct coherent mental models and evaluate their understanding (Järvelä et al., 2023). The AI's provision of Remedial Review of Basic Calculations (PostR_cluster0) also linked positively to overall SRL ($r = 0.134$), reframing what might seem like a basic request as a strategic regulatory act. Unlike clickstream data, which infers engagement via indirect proxies like page revisits (Du et al., 2023), these dialogue patterns externalize metacognitive effort, revealing how learners actively structure knowledge and strategically address weaknesses.

Examining the SRL sub-processes revealed how specific regulatory behaviors manifest through dialogue. GS emerged as the most pervasive sub-process, showing significant positive correlations with 12 of the 13 "all-positive" dialogue patterns. Pre-instruction, this included queries about surface-level prior knowledge gaps (PreQ_cluster0, r = 0.193), external learning environment obstacles (PreQ_cluster2, r = 0.193), internal learning distractions and coping strategies (PreQ_cluster3, r = 0.192), and deep learning and skill transfer (PreQ_cluster4, r = 0.168). Post-instruction, Goal Setting was associated with post-instruction strategy and focus (PostQ_cluster1, r = 0.191), targeted content clarification (PostQ_cluster2, r = 0.222), environment and distraction management (PostQ_cluster3, r = 0.192), time and environment management planning (PostQ_cluster4, r = 0.175), systematic knowledge integration (PostQ_cluster5, r = 0.293), reflection on environment and attention (PostQ_cluster6, r = 0.202), strategies for maintaining focus (PostQ_cluster7, r = 0.203), and metacognitive reflection and summary (PostQ_cluster8, r = 0.157). Additionally, the AI's remedial reviews (PostR_cluster0) correlated with GS (r = 0.174). This suggests that the act of formulating structured questions to an AI—regardless of content—is an externalization of goal-directed behavior, where learners set micro-goals to resolve uncertainty (Winne, 2020). TM was evident in four patterns: proactive gap-filling before class (PreQ_cluster0, r = 0.181), targeted clarification after class (PostQ_cluster2, r = 0.141), systematic knowledge integration (PostQ_cluster5, r = 0.204), and the use of remedial reviews (PostR_cluster0, r = 0.168). This indicates that effective time managers engage in "strategic backtracking"—investing time to address foundational issues to prevent future, more time-consuming problems (Broadbent & Poon, 2015). Finally, TS was uniquely associated with Systematic Knowledge Integration (PostQ_cluster5, r = 0.147), making this the only dialogue pattern that significantly and positively correlated with Goal Setting, Time Management, Task Strategies, and the overall SRL score. It represents a moment where the learner is not just setting goals or managing time, but is actively employing a strategy to organize the task of learning itself.

5.2.2. Inefficiency Alert Patterns

The results also identified a clear set of reactive dialogue patterns strongly associated with SRL deficits, underscoring the risks of uncalibrated AI support and highlighting the need for timely intervention. Counterintuitively, a high frequency of extensive AI scaffolding was a primary indicator of low SRL. Overall SRL was significantly and negatively correlated with AI-driven Deep Methodological Elaboration (PostR_cluster5, r = -0.230), Guidance from Concept to Application (PostR_cluster4, r = -0.220), Advanced Conceptual Preparation (PreR_cluster1, r = -0.191), Targeted Formula Deconstruction (PostR_cluster2, r = -0.169), Foundational Concept Provision (PreR_cluster0, r = -0.146), and Comparative Method Selection (PostR_cluster3, r = -0.140). Similarly, student-initiated dialogues revealing Foundational Conceptual Difficulties both before (PreQ_cluster1, r = -0.190) and after (PostQ_cluster0, r = -0.142) instruction were also robustly linked to lower overall SRL. This suggests a potential "paradox of scaffolding" or a co-regulation breakdown (Hadwin et al., 2017), where a greater volume of detailed AI explanation does not indicate effective learning but rather signals a learner who is struggling. Such patterns may be associated with cognitive overload or foster illusions of comprehension, challenging the assumption that more elaborate AI responses are always better (Mayer & Moreno, 2003; Wang & Lajoie, 2023). The strong negative correlations imply that AI responses must be carefully calibrated to learner readiness to mitigate these deficits. Future mediation analyses should test cognitive load as a potential intermediary explaining this negative link.

An examination of the SRL sub-processes reveals a cascade of regulatory failures underlying these negative patterns. Deficiencies in SE and maladaptive HS were the most pervasive issues, with all eight "all-negative" dialogue patterns showing significant negative correlations with both. For instance, student difficulties with foundational concepts (PreQ_cluster1) were linked to poor SE (r = -0.193) and HS (r = -0.168), reflecting a student's inability to accurately assess their knowledge gaps and a tendency towards inefficient help-seeking (Veenman, 2011). This pattern was amplified in the AI's most elaborate responses; the need for deep methodological elaborations (PostR_cluster5) was strongly associated with poor SE (r = -0.229) and maladaptive HS (r = -0.203), exemplifying "executive help-seeking" where the learner offloads cognitive effort to the AI rather than using it as a tool for understanding (Karabenick & Gonida, 2017). Even more foundational AI support—such as providing basic conceptual explanations before instruction (PreR_cluster0) or helping students compare solution methods (PostR_cluster3)—showed similar negative associations with SE (r = -0.173 and r = -0.172, respectively) and HS (r = -0.161 and r = -0.150, respectively). This suggests that the deficit in self-evaluation and help-seeking is not contingent on the complexity of the AI response, but rather reflects a more fundamental regulatory weakness. Furthermore, certain complex AI responses were uniquely detrimental to other regulatory functions. The provision of advanced guidance from concept to application (PostR_cluster4) was negatively associated not only with SE and HS, but also with Goal Setting (GS, r = -0.288), Task Strategies (TS, r = -0.129), and Time Management (TM, r = -0.151). Additionally, the need for deep methodological elaboration (PostR_cluster5) was negatively linked to Environment Structuring (ES, r = -0.147). These findings suggest that when a learner becomes overwhelmed to the point of needing such comprehensive guidance, their capacity to plan, strategize, manage time, and structure their learning environment is severely compromised, indicating a broader regulatory breakdown that extends beyond isolated deficits.

**5.3. Dialogue Pattern Differences Between High- and Low-SRL Groups**

These results of Mann-Whitney U Tests suggested the significant differences in dialogue patterns found between learners with high- and low-SRL proficiency. The results of this group comparison complement the preceding correlational analysis, providing convergent evidence that specific dialogue patterns can serve as valid indicators of a learner's SRL status.

A key finding from this analysis is that learners with lower demonstrated SRL proficiency engaged significantly more frequently in dialogues concerning fundamental conceptual difficulties. This was observed both before instruction (PreQ_cluster1) and after instruction (PostQ_cluster0). This observation aligns with established research in self-regulated learning, which posits that learners with regulatory deficits often struggle with accurate self-assessment and are more likely to possess unresolved knowledge gaps (Veenman, 2011). The dialogues appear to function as an externalization of these internal regulatory challenges. Whereas traditional data sources like clickstream activity may register user engagement without capturing its underlying purpose, these specific question patterns provide a more direct window into the learner's cognitive and metacognitive state, signaling an ongoing struggle with core content that is characteristic of lower SRL.

In addition to student-initiated dialogues, the patterns of AI responses triggered by the two groups also differed significantly, offering further insight into the nature of SRL deficiencies. The finding that the low-SRL group prompted a higher frequency of scaffolding-intensive AI responses—from foundational concept provision (PreR_cluster0) to deep methodological elaborations (PostR_cluster5)—presents a "paradox of scaffolding." Rather than indicating effective use of the learning tool, a high rate of elicited support appears to signal a dependency that stems from an inability to self-regulate effectively. This resonates with theories of co-regulation and cognitive load, which suggest that while scaffolding is intended to be supportive, its excessive use can point to a breakdown in the learner's ability to monitor their own learning, set appropriate goals, and select strategies, leading them to offload cognitive responsibility to the external agent (Hadwin et al., 2017; Mayer & Moreno, 2003).

Taken together, the results of the group comparison provide convergent evidence for the validity of using dialogue analytics to assess SRL. The combination of reactive, student-initiated questioning about core concepts and a corresponding high frequency of elicited AI scaffolding forms a distinct behavioral signature for learners with lower SRL. Crucially, the fact that the dialogue patterns identified as significant in this group-level analysis are the same ones that showed strong negative correlations with SRL in the preceding analysis provides robust, triangulated evidence for their utility as indicators. This moves beyond mere association to establish these patterns as reliable markers that can differentiate between high and low levels of self-regulation in this learning context.

**5.4. A Framework for Tracing SRL in Dialogue-Based Human-AI Collaborative Learning**

The findings from this study are operationalized in the Dialogue-Based Human-AI Self-Regulated Learning (DHASRL) framework (Table 4), which extends the conceptual model, HASRL, proposed by Järvelä et al. (2023). While the HASRL model provides a foundational blueprint for how human and AI regulation can be shared, our DHASRL framework provides the empirical and operational detail necessary for its implementation in dialogue-based learning environments.

Specifically, we extend the HASRL model in three critical ways. First, we give concrete form to the abstract concepts of Signals and Traces by identifying specific, empirically-derived dialogue patterns that serve as reliable indicators of students' underlying regulatory processes. Second, we provide the crucial interpretative layer for the AI's Detect and Diagnose functions. Our framework moves beyond simple pattern detection to a pedagogically meaningful diagnosis, classifying patterns into 'Proactive Indicators' (signaling effective regulation) and 'Inefficiency Alerts' (warning of breakdowns). This diagnostic capability directly informs the AI's third function, Act, by providing a clear basis for when to reinforce and when to intervene. Finally, by grounding these patterns in Zimmerman's (2002) cyclical phases (such as Goal Setting in Table 4.), our framework creates a robust link between the AI's real-time diagnosis and the student's internal SRL state, thereby operationalizing the entire Human-AI feedback loop central to the HASRL model.

**Table 4.**

*The Dialogue-Based Human-AI Self-Regulated Learning Tracing Framework.*

| Indicator Category | Interpretation | Associated Dialogue Clusters | Primary SRL Components |
|---|---|---|---|
| Proactive Regulation Patterns (High frequency is desirable) | | | |
| 1. Proactive Planning & Preparation | Student sets the stage for future learning by checking prerequisites, planning time, and managing their environment and internal state. | PreQ_cluster0, PreQ_cluster2, PreQ_cluster3, PostQ_cluster4 | GS, TM |
| 2. Strategic Knowledge Integration & Remediation | Student actively works to clarify, connect, and remediate their understanding of course material through targeted and summative questions. | PreQ_cluster4, PostQ_cluster2, PostQ_cluster5, PostQ_cluster8, PostR_cluster0 | Overall SRL, GS, TM, TS |
| 3. Metacognitive Management of Learning Process | Student monitors and regulates the conditions of learning, such as their attention, focus, and study environment, separate from the content itself. | PostQ_cluster1, PostQ_cluster3, PostQ_cluster6, PostQ_cluster7 | Overall SRL, GS, TM |
| Inefficiency Alert Patterns (High frequency is a warning signal) | | | |
| 4. Reactive Remediation of Foundational Knowledge | Student is trapped in a cycle of asking about basic concepts, indicating a failure to prepare adequately or learn effectively during instruction. | PreQ_cluster1, PostQ_cluster0 | Overall SRL, SE, HS, ES |
| 5. Cognitive Independence or Overload via | Student triggers a cascade of detailed AI explanations which, due to a lack of prerequisite knowledge, | PreR_cluster0, PreR_cluster1, PostR_cluster2, PostR_cluster3, | Overall SRL, GS, SE, HS, TM, TS, ES |

| | | |
|---|---|---|
| Indiscriminate Help-Seeking | likely induce overload rather than understanding. | PostR_cluster4, PostR_cluster5 |

*Note*: Primary SRL Components include SRL overall (Overall SRL) performance and six subprocess: goal setting (GS), environment structuring (ES), task strategies (TS), time management (TM), help-seeking (HS), and self-evaluation (SE).

The first, Proactive Regulation Indicators, encompasses three distinct patterns of behavior that are desirable and indicative of effective self-regulation. A high frequency of dialogue falling into these categories—such as a student asking planning questions (Proactive Planning & Preparation) or summative questions (Strategic Knowledge Integration)—signals a healthy, agentic learning process. These are the behaviors that an ideal HAICL system would seek to encourage. These patterns reflect a learner actively engaging in the forethought and self-reflection phases of SRL (Zimmerman, 2002), setting clear goals and proactively monitoring their progress.

Secondly, the Inefficiency Alert Patterns identify two recurring patterns that signal a breakdown in the learning process. A high frequency of dialogue in these categories serves as a warning signal for intervention. For example, the emergence of the Reactive Remediation pattern, where a student repeatedly asks about foundational concepts, is a strong indicator of a systemic failure in their learning cycle. This pattern reveals a breakdown in the cyclical process of SRL, as the learner continually attempts to remediate foundational knowledge gaps rather than engaging in deeper learning. Likewise, the Cognitive Independence or Overload pattern, identified by the student's passive reception of numerous detailed AI explanations, suggests that the student is overwhelmed (Wang & Lajoie, 2023) or less autonomy (Zhai et al., 2024). This reflects a failure in metacognitive monitoring and self-evaluation, where the learner is unable to effectively process complex information, self-evaluation or seek appropriate assistance. In practice, an AI system could implement this framework by classifying dialogue in real-time and monitoring the frequency of these five patterns. The identification of a specific pattern is the diagnostic step; the crucial subsequent step, discussed next, is determining the appropriate pedagogical response.

**Figure 2.**

*The Dialogue-Based Human-AI Self-Regulated Learning Relationship Framework.*

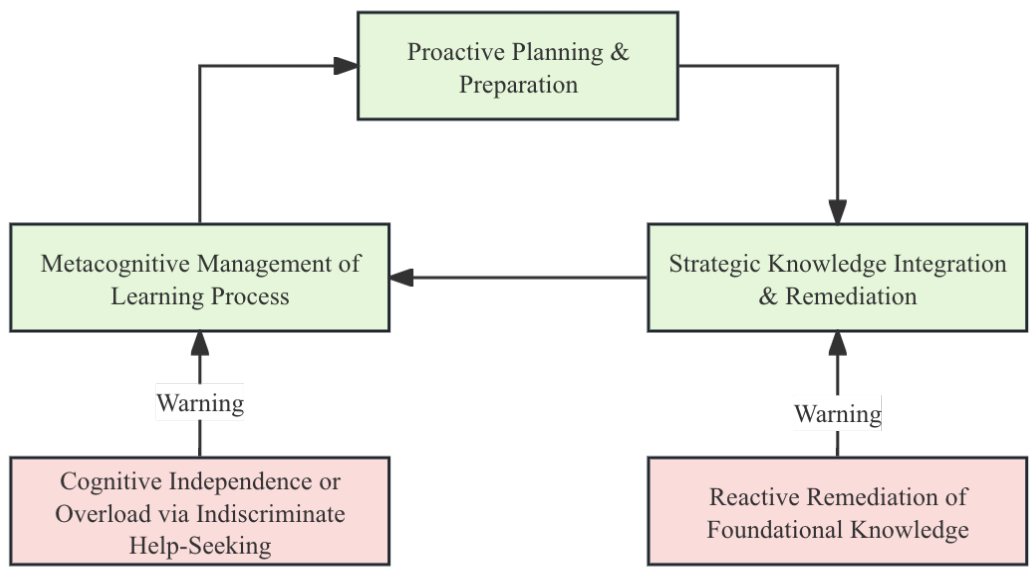

Shown in Figure 2, these five indicators are not isolated but exist in a dynamic, oppositional relationship that illustrates the flow of SRL in a HAICL environment. The three Proactive Regulation Patterns represent a healthy, functional SRL cycle aligned with Zimmerman's (2002) model. This cycle is initiated by Proactive Planning & Preparation (forethought), which enables effective Strategic Knowledge Integration (performance), and is guided by Metacognitive Management (self-reflection and monitoring) to inform future cycles. The two Inefficiency Alert Patterns function as direct negative counterparts to the initiation and guidance of this cycle. The emergence of Reactive Remediation is a symptom of failed Proactive Planning; a student is forced to reactively fix foundational gaps because they did not proactively prepare. Similarly, the appearance of Cognitive Overload signals a breakdown in Metacognitive Management, where a failure to monitor one's own comprehension leads to counterproductive help-seeking. In this model, effective Strategic Knowledge Integration is the desired outcome, which is only achievable when the proactive planning and metacognitive management cycles are intact.

### 5.5. Implications for Pedagogy and Practice

These findings advance HAICL by emphasizing SRL-sensitive systems. Educators can integrate DHASRL into Gen AI-based learning system for real-time monitoring to trigger reflection prompts (e.g., "How does this integrate with prior knowledge?"), or inefficiency alerts for guided interventions, (e.g., it could pause the elaborate explanation and pivot to a different strategy). In humanities examples, like creative writing, proactive patterns could adapt AI to scaffold task integration, fostering SRL by prompting "Reflect on narrative applications." The DHASRL based AI instructors oversee AI outputs to ensure beneficial and ethical content (e.g., a judgement agent for detecting the output of Gen AI, used to confirm that the content is beneficial for students' SRL, Qian et al., 2026). System designers gain actionable criteria to calibrate responses, balancing support with autonomy to enhance engagement and equity.

Furthermore, our findings also provided implications for teachers. Teachers should require students to submit their dialogue logs with the AI to assess whether their SRL skills have been affected. Teachers should remind students to avoid relying excessively on AI for detailed explanations of knowledge, instead encouraging them to learn to use AI to assist in managing their studies and improving their metacognition.

**5.6. Limitations**

Several limitations of this study warrant consideration. First, regarding the validity of our identified patterns, while the group comparison results substantiated the "Inefficiency Alert Patterns" as indicators of SRL challenges, the evidence for "Proactive Regulation Patterns" was less conclusive. This, combined with the generally low correlation coefficients, suggests that a student's SRL process is a holistic phenomenon influenced by unmeasured variables, such as motivation or prior knowledge, that are not fully captured by dialogue data alone. Second, our findings' generalizability is constrained by the homogeneous sample of finance undergraduates in a statistics context; future research is needed to validate these patterns across diverse disciplines and educational levels, such as K-12 or non-STEM fields. Third, the reliance on a large language model for analysis introduces the potential for inherent biases (e.g., cultural prejudices) to distort dialogue interpretation (Kasneci et al., 2023).

Looking forward, a more robust research program is required. Longitudinal studies are needed to track how SRL trajectories evolve and to calibrate diagnostic thresholds (high frequent criteria) over time. Most critically, future work should employ rigorous randomized controlled trials (RCTs) to firmly establish the causal efficacy of the DHASRL intervention and explore its potential for scalable application in real-world educational settings.

## 6. Conclusion

This study demonstrates that student-AI dialogues provide a rich and observable source of data for understanding student SRL. By establishing significant correlations between specific dialogue patterns and SRL assessments, and by identifying distinct conversational behaviors that differentiate high- and low-SRL student groups, our findings confirm that dialogue features can serve as reliable behavioral indicators of SRL proficiency. On a theoretical level, this work provides a concrete operationalization of socially shared regulation of learning within HAICL, demonstrating that the dialogue itself is the observable venue where the co-regulatory process between a student and an AI unfolds. Building on this evidence, we introduce a novel exploratory framework for the Dialogue-Based Human-AI Self-Regulated Learning Framework (DHASRL). This framework offers a methodology to track and even provide early warnings for students' SRL capabilities based on their conversational patterns within a Human-AI Collaborative Learning context. For practice, this approach lays the groundwork for next-generation generative AI-based pedagogical systems capable of real-time student monitoring and targeted intervention. For research, our work contributes a new methodological layer to learning analytics, complementing traditional clickstream data with nuanced conversational indicators. This integration promises more accurate and timely models for identifying students who require support. Ultimately, this research moves beyond static self-reports, offering a scalable and dynamic approach to assessing and fostering the critical learning-to-learn skills essential in the age of AI.

Appendix A.

A.1. Approach of Dialogue Patterns Identification and Alignment

To identify recurring, meaningful patterns within the vast amount of unstructured text data, we employed a large language model based embedding cluster pipeline.

Sentence Embedding: First, all student questions and AI responses were transformed into high-dimensional numerical vectors. This was achieved using the Alibaba nlp_gte_sentence-embedding model, which is optimized for understanding the semantic content text.

Clustering: To group semantically similar utterances, a two-step clustering approach was used. The UMAP (Uniform Manifold Approximation and Projection) algorithm was first applied to the sentence embeddings to reduce their dimensionality while preserving the essential data structure. Subsequently, the K-Means clustering algorithm was applied to this reduced-dimension data to group the utterances into distinct clusters. Each resulting cluster represents a recurring "dialogue patterns"—a typical type of question or response (e.g., a question about foundational concepts, an AI response providing a deep conceptual explanation).

**Table A.1**

*Dialogue Feature Identification Pipeline Overview*

| Stage | Method | Input | Output |
| --- | --- | --- | --- |
| Segmentation | Manual categorization | 421 dialogue logs | Pre/post questions & answers |
| Embedding | Alibaba nlp_gte_sentence-embedding-v4 | Text segments | 1024-dimensional vectors |
| Reduction | UMAP | High-dimensional vectors | Reduced representations |
| Clustering | K-means or HDBSCAN | Reduced vectors | Thematic clusters |
| Interpretation | Qualitative analysis | Clusters | 22 conversational features |

A progressive random search strategy was employed to determine the optimal hyperparameters for both the UMAP and K-means stages. For UMAP, parameters were optimized across ranges for n_neighbors (5-50), min_dist (0.01-0.5), and n_components (2-30). For K-means, the number of clusters was evaluated within a range of 2 to 10. The optimization process consisted of: (1) a coarse parameter search across 100 iterations to identify promising parameter regions, and (2) a fine-tuning phase around the top 5% of results through 30 additional iterations per configuration.

To objectively evaluate the quality of the resulting clusters, we used a Composite Score combining three normalized cluster validity indices: the Silhouette coefficient, the Calinski-Harabasz index, and the Davies-Bouldin index. These indices quantify different aspects of cluster quality, such as separation and compactness. A weighted combination of these indices allowed us to balance these considerations, resulting in a composite score:

*Composite Score*
$$= 0.5 \times Silhouette\_norm + 0.3 \times CH\_norm + 0.2 \times (1 - DB\_norm)$$

The weights were assigned based on the relative importance of each metric in reflecting overall cluster quality, with the Silhouette coefficient given the highest weight due to its sensitivity to both cluster cohesion and separation. The metrics were normalized to a 0-1 scale before combination. We set the target as Composite Score and carried out the hyperparameters search strategy.

The optimal hyperparameters for this process, detailed in Table A.2, were determined through a systematic search strategy, yielding strong cluster separation, with silhouette scores ranging from 0.46 to 0.58 (the best Composite Score for student questions: 0.995; the best Composite Score for AI responses: 0.952).

**Table A.2.**

*Optimal Clustering Parameters*

| Parameter | Pre-Questions | Post-Questions | Pre-Answers | Post-Answers |
|---|---|---|---|---|
| UMAP | | | | |
| n_neighbors | 5 | 10 | 50 | 20 |
| min_dist | 0.2 | 0.3 | 0.01 | 0.2 |
| n_components | 5 | 2 | 2 | 30 |
| metric | euclidean | euclidean | euclidean | cosine |
| Clustering | | | | |
| algorithm | K-means | K-means | K-means | K-means |
| n_clusters | 5 | 9 | 2 | 6 |
| Performance | | | | |
| Silhouette Score | 0.5066 | 0.5806 | 0.5383 | 0.4599 |
| CH Index | 195.80 | 1167.56 | 737.20 | 458.02 |
| DB Index | 0.6825 | 0.4795 | 0.6987 | 0.8154 |

With the optimal hyperparameters, our embedding clustering based on the large language model shows 22 distinct clusters. The semantic coherence of the resulting clusters were verified through qualitative examination of cluster contents by two independent raters. Inter-rater agreement was assessed using Cohen's kappa (κ), a statistical measure of agreement that accounts for the possibility of agreement occurring by chance. κ = 0.87 indicates substantial agreement between the raters, suggesting a high degree of consistency in their interpretation of the cluster contents. Consequently, the naming and interpretation as dialogue patterns of these 22 clusters are shown in Table 1.

After identifying the dialogue patterns, we needed to quantify each student's propensity to exhibit them. This was achieved by calculating the cosine similarity [0,1] between the vector of each individual student utterance and the centroid vector of each identified cluster. This similarity represents the degree to which a specific interaction reflects a given dialogue feature, serving as a pattern alignment scores.